\documentclass[epj]{webofc}
\usepackage[varg]{txfonts}   
\woctitle{MESON2014 - the 13$^\textrm{th}$ International Workshop on Meson Production, Properties and Interaction}
\begin{document}
\selectlanguage{english}
\title{Signature of an $h_1$ state from $J/\psi  \to \eta K^{*0}\bar{K}^{*0}$ and theoretical description of the $Z_c(3900)$ and $Z_c(4020)$ as $D \bar D^*$ and $D^* \bar D^*$ molecular states}

\author{E. Oset\inst{1}\fnsep\thanks{\email{oset@ific.uv.es}} \and
        Ju Jun Xie\inst{2} \and
        M. Albaladejo\inst{1} \and F. Aceti\inst{1} \and M.Bayar\inst{3} \and
J.M.Dias\inst{1,4} \and A. Martinez Torres\inst{4} \and K. Khemchandani\inst{4}
\and M. Nielsen\inst{4} \and F. Navarra\inst{4} \and X.L. Ren\inst{5} \and L.S. Geng\inst{5}
\and Jie Meng\inst{5} \and W. H. Liang\inst{6}
}

\institute{Departamento de
F\'{\i}sica Te\'orica and IFIC, Centro Mixto Universidad de
Valencia-CSIC Institutos de Investigaci\'on de Paterna, Aptdo.
22085, 46071 Valencia, Spain 
\and
           Institute of Modern Physics, Chinese Academy of
Sciences, Lanzhou 730000, China
\and Department of Physics, Kocaeli University, 41380 Izmit, Turkey
\and
      Instituto de F\'isica, Universidade de S\~ao Paulo,
C.P. 66318, 05389-970 S\~ao Paulo, SP, Brazil
 \and School of Physics and
Nuclear Energy Engineering \& International Research Center for Nuclei and Particles in the Cosmos, Beihang University, Beijing 100191, China    
\and Department of Physics, Guangxi Normal University, Guilin, 541004, P. R. China     }

\abstract{In this talk we address two topics: The first one is an empirical explanation in terms of a new state $h_1$ of the peak in the $K^{*0}\bar{K}^{*0}$ invariant mass distribution close to threshold of this channel in the  $J/\psi  \to \eta K^{*0}\bar{K}^{*0}$ decay.  The second one is a theoretical description of the isospin $I=1$ $Z_c(3900)$ and $Z_c(4020)$ states in terms of molecular states of $D \bar D^*+ cc$ and $D^* \bar D^*$.
 
}
\maketitle
\section{Introduction}
\label{intro}

  In this talk we will address two topics: The explanation of the enhancement of the $K^{*0}\bar{K}^{*0}$ invariant mass distribution close to threshold in the  $J/\psi  \to \eta K^{*0}\bar{K}^{*0}$ decay in terms of a new state $h_1$ and the theoretical description of the $Z_c(3900)$ and $Z_c(4020)$ in terms of molecular states of $D \bar D^*$ and $D^* \bar D^*$.

\section{Experimental support for a new $h_1(1830)$ state}
\label{sec-1}
\begin{figure}[t!]\centering
\includegraphics[width=0.50\textwidth]{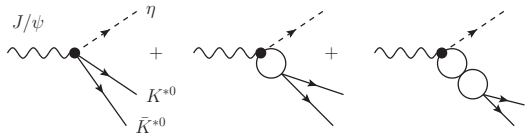}
\caption{Diagrammatic representation of the $J/\psi \to \eta
K^{*0}\bar{K}^{*0}$ decay. \label{feydgm}}
\end{figure}

The decay $J/\psi \to \eta K^{*0}\bar{K}^{*0}$ was measured
by the BES Collaboration~\cite{BESdata} 
searching for the $Y(2175)$ resonance, however, no clear enhancement in the $K^{*0}\bar{K}^{*0}$ mass distribution was found near $2.175$ GeV.  Instead an enhancement in the $K^{*0}\bar{K}^{*0}$ mass distribution was found that passed unnoticed.  In \cite{xiemiguel} this fact was used to claim support for an $h_1$ resonance that was predicted in \cite{gengvec} as dynamically generated from the interaction of vector mesons. The $h_1$ state has quantum numbers $0^-(1^{+-})$, and, since the $J/\psi$ and the $\eta$ mesons have quantum numbers $0^-(1^{--})$ and $0^+(0^{-+})$, respectively, the decay $J/\psi \to \eta K^{*0}\bar{K}^{*0}$ constitutes the ideal reaction to look for an $h_1$ state, coupling to an $s$--wave $K^* \bar{K^*}$ pair. In Fig. \ref{feydgm} we show diagrammatically the process. The $K^{*0}\bar{K}^{*0}$ state is produced and the two $K^*$ interact. This interaction is the one that leads to the $h_1$ resonance in \cite{gengvec}. It is interesting to note that this state is very peculiar. Because of its negative C-parity, the  $h_1$
state cannot couple to two vector mesons of the type $\rho, \omega, \phi$. Also because of its positive parity it cannot couple to two pseudoscalar mesons, which would have to be in L=1 to match the spin of the $h_1$ and hence would have negative parity. The  $K^*\bar{K}^*$ is then a unique state for both the construction and decay of this $h_1$ state.

The scattering matrix for the $K^*\bar{K}^*$ interaction is given in terms  of a potential $V$ by 
\begin{equation}\label{eq:tmat}
t = v+v\widetilde{G}t = v ( 1+ \widetilde{G} t) =(1-v\widetilde{G})^{-1} v = (v^{-1} - \widetilde{G})^{-1}~,
\end{equation}
where $\widetilde{G}(M_\text{inv}^2)$ is the loop function for the $K^* \bar{K}^*$ pair. This function is divergent, and it can be regularized both with a cutoff  or with
dimensional regularization in terms of a subtraction
constant. On the other hand the full amplitude for the process $J/\psi \to \eta K^* \bar{K}^*$ can be written, according to the diagrams in Fig.~\ref{feydgm}, as:
\begin{equation}
t_P = V_P \left(1 + \widetilde{G}(M_\text{inv}^2) t(M_\text{inv}^2) \right) = V_P \frac{t(M_\text{inv}^2)}{v(M_\text{inv}^2)}~,
\end{equation}
where the last equality follows from Eq.~\eqref{eq:tmat}.

Our strategy to interpret the data of \cite{BESdata} is to use the potential $V$ derived in \cite{gengvec} and fit the data of \cite{BESdata} by changing the subtraction constant in the $\widetilde{G}$ function in dimensional regularization $a(\mu)$, where $\mu$ is a regularization scale, taken here as $\mu = 1\ \text{GeV}$.

\begin{figure}[t!]\centering
\includegraphics[width=0.45\textwidth]{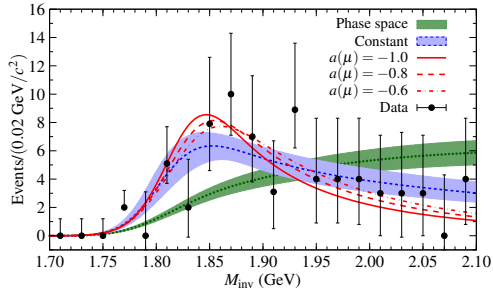}
\caption{(Color online) The $K^{*0}\bar{K}^{*0}$ invariant mass
spectrum of $J/\psi \to \eta K^{*0}\bar{K}^{*0}$ decay. The data points are taken from
Ref.~\cite{BESdata}. The different lines represent the output for the different approaches considered. The short-dashed line and the associated error band (light blue) represent the results of the constant potential. The (red) solid, long-dashed and dot-dashed lines represent the results for the  vector exchange potential of \cite{gengvec}  with $a(\mu)=-1.0$, $-0.8$ and $-0.6$, respectively. Finally, the (dark green) dotted line, and the associated error band (green) is the prediction for phase space alone.\label{fig:dgdm}}
\end{figure}

 As seen in figure \ref{fig:dgdm}  a good agreement with the data is found for reasonable values of the parameter $a(\mu)$. The figure also shows the results with pure phase space normalized to the same area and we see that the disagreement with the data is quite obvious. 

\begin{figure}[t!]\centering
\includegraphics[width=0.40\textwidth]{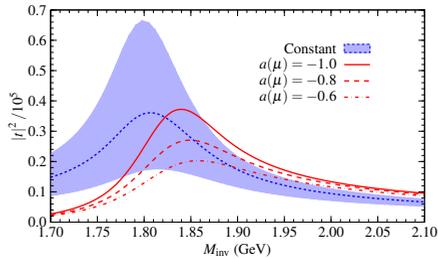}
\caption{(Color online) The modulus squared of the amplitude, $\left\lvert t \right\rvert^2$, for
$K^*\bar{K^*} \to K^*\bar{K^*}$, for the different approaches considered in this work. The notation of the lines as in Fig.~\ref{fig:dgdm}.\label{fig:ts}}
\end{figure}

In order to see that with this parameter $a(\mu)$ one gets a resonance, in figure \ref{fig:ts} we show the magnitude of $|t|^2$ and we see indeed that it has a resonant shape. We also perform an alternative fit to the data which could be carried on by someone not knowing the potential of \cite{gengvec}. In this case we can take a constant potential and fit it to the data. The results obtained are similar and the conclusion the same. Altogether we find from this analysis that there is a resonance $h_1$ with a mass $M_{h_1} = 1830 \pm  20\ \text{MeV}$ and $\Gamma_{h_1} = 110 \pm 10\ \text{MeV}$, respectively.

\begin{figure}[t!]
  \centering
  \includegraphics[width=5cm]{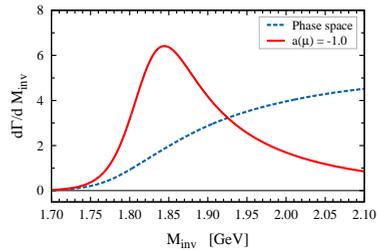}\\
  \caption{(Color online). The $K^{*+}K^{*-}$ invariant mass spectrum of $\eta_c(2S)\rightarrow\phi K^{*}\bar{K}^{*}$ decay. The red solid line represents the results obtained with
   $a(\mu)=-1.0$, and the blue dashed line is the  phase space.}
  \label{Fig:Fig4}
\end{figure}

We think that these data offer enough evidence for this new state.  In \cite{xiulei} we have made predictions for two more reactions $\eta_c\rightarrow \phi K^*\bar{K}^*$, or $\eta_c(2S)\rightarrow \phi K^*\bar{K}^*$, that should help establish this resonance on firmer grounds. In Fig. \ref{Fig:Fig4} we show the predictions for $\eta_c(2S)\rightarrow \phi K^*\bar{K}^*$ comparing the results with the $h_1$ resonance and pure phase space. The differences are striking and we encourage such a measurement to give further support to our claims based on the $J/\psi \to \eta K^{*0}\bar{K}^{*0}$ decay. 

  In the charm sector there is also a related $h_1$ hidden charm state stemming from the vector vector interaction in \cite{raquelxyz} at 3945 MeV, which is also found
in \cite{juan1}.  In \cite{weih} a reaction is also proposed to find this state, looking at the invariant mass distribution of $D^* \bar D^*$ in the $X(4660) \to \eta  D^* \bar D^*$ decay. 

\section{theoretical description of the $Z_c(3900)$ and $Z_c(4020)$ as $D \bar D^*$ and $D^* \bar D^*$ molecular states}

One of the interesting recent surprises in hadronic physics has been the discovery of states that challenge the standard $q \bar q$ structure of the mesons. In this sense, charmonium states of isospin I=1 are of this type since $c \bar c $ has necessarily I=0. One state called $Z_c(3900)$
has been reported by the BESIII collaboration \cite{bes3900} and reconfirmed in other experiments. Similarly, in \cite{bes4020} another $Z_c(4020)$ has been reported, also reconfirmed in other experiments. It is unclear whether this state has something to do with the peak in the   $D^* \bar D^*$ invariant mass distribution at 4025 MeV in the $e^+ e^- \to (D^* \bar D^*)^{\pm}  \pi^{\mp}$  reaction studied in \cite{Ablikim:2013emm}. Indeed, it was shown in \cite{alberto} that this peak could be interpreted in terms of a resonance at much lower energy that was predicted in \cite{raquelxyz} from the $D^* \bar D^*$ interaction.

\begin{figure}[t!]
\centering
\includegraphics[width=0.30\textwidth]{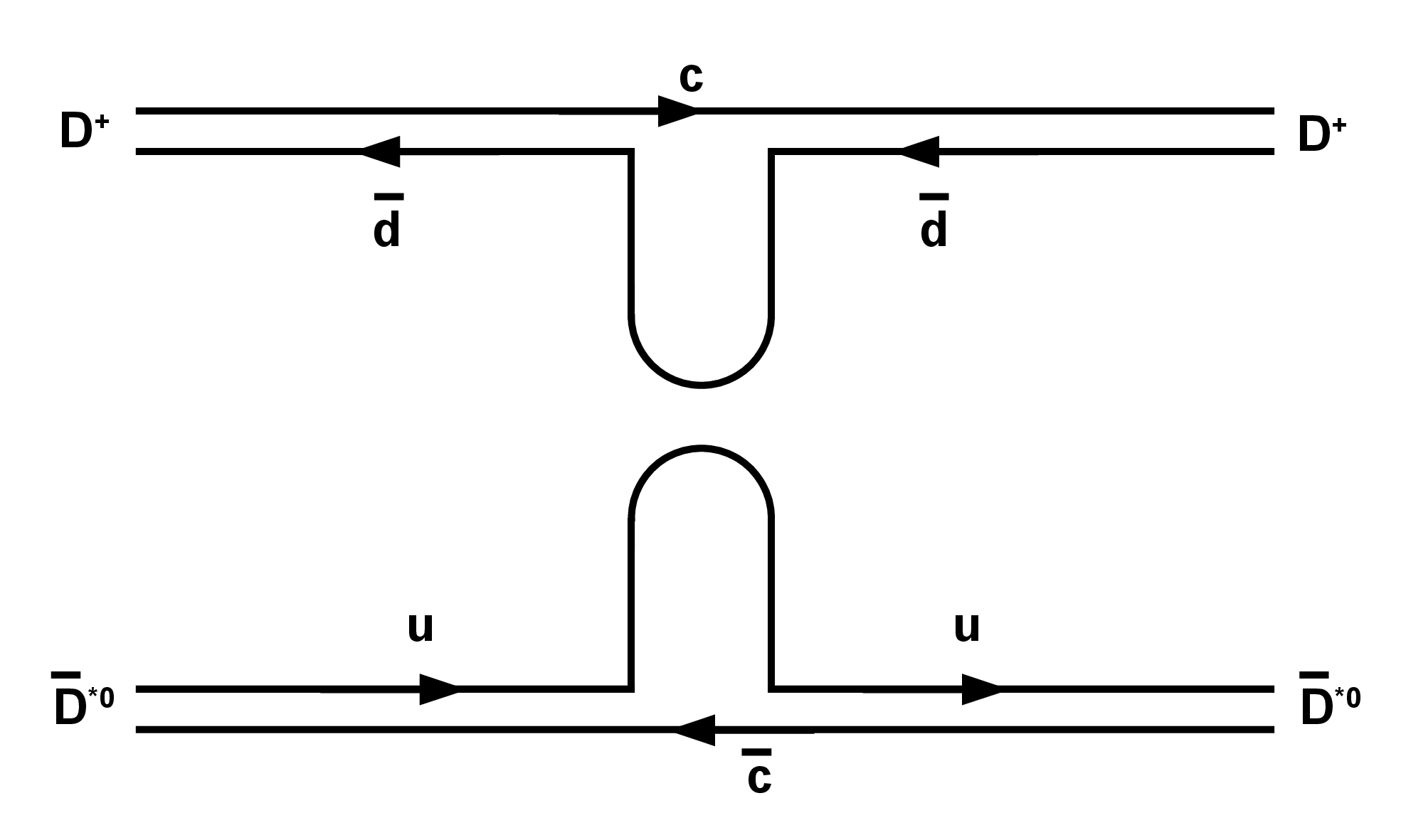}
\caption{Feynman diagram depicting the OZI forbidden exchange of a light $q\bar{q}$ pair.}
\label{fig:OZI}
\end{figure}

In \cite{ddstar,dstardstar} we have looked into the possibility of creating states from the interaction of $D \bar D^* +cc$ and $D^* \bar D^*$ with isospin 1. In order to study the interaction of pseudoscalar of vector mesons we use the local hidden gauge approach \cite{hidden1}, which generalizes the chiral Lagrangians to include vector mesons.  The approach produces the interaction by means of the exchange or vector mesons and several states can be produced with I=0 in the charm sector \cite{raquelxyz}, or in the bottom sector \cite{xiaozpi}.
Yet, the I=1 states are more difficult to obtain. The reason can be devised by looking at fig. \ref{fig:OZI}. We can see that if we try to exchange a light $q \bar q$ state it is OZI forbidden since we would need to exchange a $d \bar d $ state for the upper line and the lower line demands a $u \bar u$ one.  In view of this one has to exchange a heavy vector $c \bar c$, but one pays a penalty in this mechanism since the propagator of the heavy vector is much suppressed. This makes the interaction weaker and the possibilities of binding smaller. What we say about the exchange of a light vector can also be said about the exchange of pseudoscalar mesons and in \cite{ddstar,dstardstar} it was shown that in the limit of equal masses for the pseudoscalar mesons, the contribution of the $\pi, \eta, \eta'$ together vanished. In the case when different masses are taken into account then the cancellation is partial but the mechanism is also quite suppressed.

\begin{figure}[t!]
\centering
\includegraphics[width=0.37\textwidth]{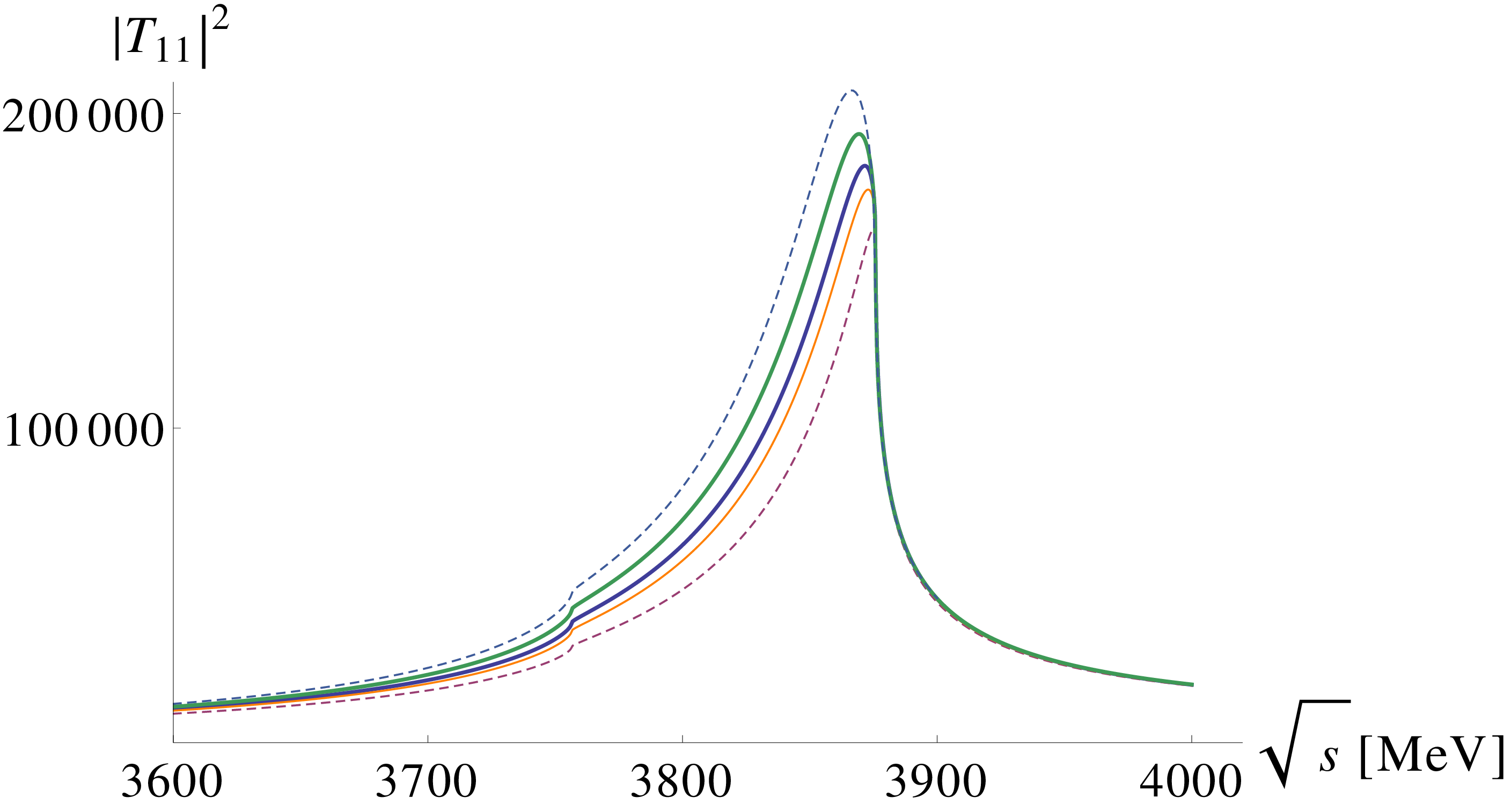}
\caption{$|T|^2$ as a function of $\sqrt{s}$ for values of the cutoff $q_{max}$ equal to $850$, $800$, $770$, $750$ and $700$ MeV. The peak moves to the left as the cutoff increases.}
\label{fig:cutoffs}
\end{figure}

\begin{figure}[t!]
\centering
\includegraphics[scale=0.30]{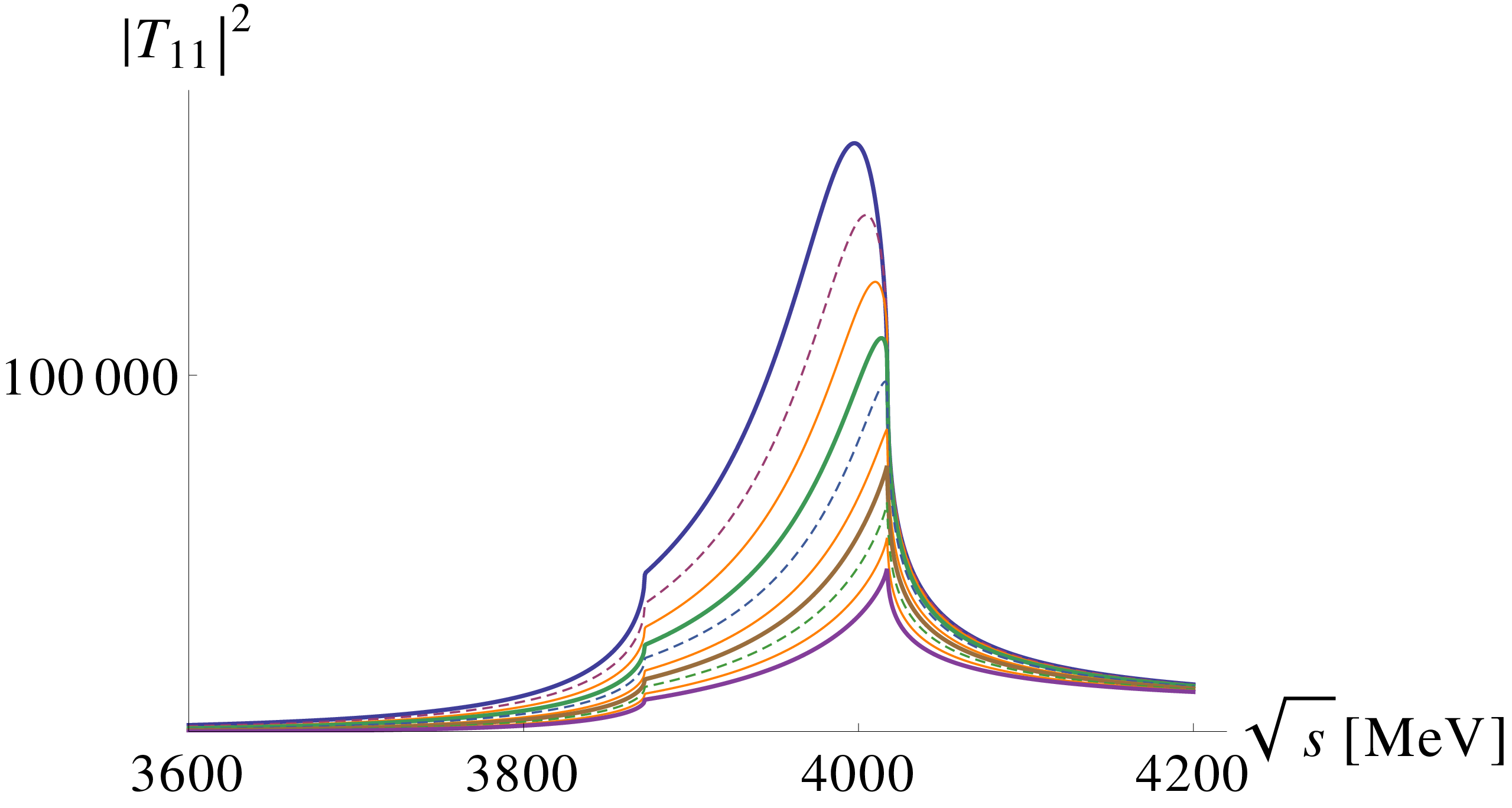}
\caption{$|T_{11}|^2$ as a function of $\sqrt{s}$, for different values of the cutoff $q_{max}$. From up down, $q_{max}=960,\,900,\,850,\,800,\,750,\,700,\,650,\,600,\,550,\,500$ MeV.}
\label{fig:cutoff}
\end{figure}

 In view of this, in \cite{ddstar,dstardstar} we also studied the exchange of two pions, either correlated to give the $f_0(500)$ or $\sigma$ meson through their interaction, or uncorrelated. They also were found to be relatively small. Yet, the strength of the heavy vector exchange, even if very much weakened, is enough to create some structure, pole or cusp, depending on the strength of the interaction which is governed by the cut off of the G function. This we can see in Figs. \ref{fig:cutoffs} and \ref{fig:cutoff} respectively for the case of the interaction of  $D \bar D^* +cc$ and $D^* \bar D^*$.
We can see in the figures that indeed the strength of $|T|^2$ accumulates around the thresholds of 
 $D \bar D^* $ and $D^* \bar D^*$ respectively, and one could claim that a resonant state is obtained close to the threshold of these two channels, very close to what is observed in the experiment.  In our case the width comes because the calculations are done in coupled channels and there are open channels, $\eta_c \rho$ and $\pi J/\psi$ in the case of the  $D \bar D^* +cc$ states and $\rho J/\psi$ in the case of $D^* \bar D^*$ states. The width is about 50 MeV for the case of the  $D \bar D^* +cc$ state and about 100 MeV for the case of the $D^* \bar D^*$ state, in qualitative agreement with experiment. 

\section{Conclusion}
We have reported work on two topics. In the first one we showed how the data of the $J/\psi \to \eta K^{*0}\bar{K}^{*0}$ decay, by looking at the invariant mass distribution of the  $K^{*0}\bar{K}^{*0}$, provided some evidence for a new $h_1(1830)$ state which is not catalogued in the PDG. We also suggested performing a measurement of two more reactions, $\eta_c\rightarrow \phi K^*\bar{K}^*$, or $\eta_c(2S)\rightarrow \phi K^*\bar{K}^*$, that should help establish this resonance on firmer grounds. In the second part we explored the possibility of forming molecular states of 
$D \bar D^* +cc$ and $D^* \bar D^*$. We stressed the observation that from the point of view of meson exchange, the exchange of light vector mesons or pseudoscalar mesons is OZI forbidden. Then we explored the exchange of heavy vectors and two pion exchange, correlated and uncorrelated. We still found the exchange of vector mesons bigger than the two pion exchange, and enough to create some molecular structure very close to the threshold of the $D \bar D^* +cc$ and $D^* \bar D^*$ channels. We also included other pseudoscalar-vector and vector-vector states in a coupled channels formalism, which were open for the decay of the states found and they gave rise to a width in qualitative agreement with experiment, thus giving support for the $Z_c(3900)$ and $Z_c(4020)$ as weakly bound molecular states of $D \bar D^* +cc$ and $D^* \bar D^*$ respectively.

\begin{acknowledgement}
This work is partly supported by the Spanish Ministerio de Economia y Competitividad and European FEDER funds under the contract number
FIS2011-28853-C02-01, and the Generalitat Valenciana in the program PrometeoII, 2014/068. J. M. Dias acknowledges the Brazilian Funding Agency FAPESP for support. We acknowledge the support of the European Community-Research Infrastructure Integrating Activity Study of Strongly Interacting Matter ( HadronPhysics3, Grant 
no. 283286) under the Seventh Framework Programme of EU and the National
Natural Science Foundation of China under Grant No. 11165005 and Guangxi 201203YB017.
\end{acknowledgement}
%
%
%

\end{document}